\documentclass[twocolumn]{emulateapj}
\usepackage{graphicx}
\usepackage{dcolumn}

\graphicspath{{./Users/nathansecrest/Documents/Papers/ngc4178agn/}}

\def\arcsec{\hbox{$^{\prime\prime}$}}
\def\deg{\hbox{$^\circ$}}

\begin{document}

\title{The {\it Chandra} View of NGC 4178: The Lowest Mass Black Hole in a Bulgeless Disk Galaxy?}

\author{N.~J.~Secrest\altaffilmark{1}, S.~Satyapal\altaffilmark{1}, M.~Gliozzi\altaffilmark{1}, C.~C.~Cheung\altaffilmark{2}, A.~C.~Seth\altaffilmark{3}, \& T.~B\"{o}ker\altaffilmark{4}}

\altaffiltext{1}{George Mason University, Department of Physics \& Astronomy, MS 3F3, 4400 University Drive, Fairfax, VA 22030, USA}

\altaffiltext{2}{National Research Council Research Associate, National Academy of Sciences, Washington, DC 20001, resident at Naval Research Laboratory, Washington, DC 20375, USA}

\altaffiltext{3}{Department of Physics and Astronomy, University of Utah, Salt Lake City, UT 84112, USA}

\altaffiltext{4}{ESA/ESTEC, Keplerlann 1, 2200 AG Noordwijk, Netherlands}

\begin{abstract}

Using high resolution {\it Chandra} data, we report the presence of a weak X-ray point source coincident with the nucleus of NGC 4178, a late-type bulgeless disk galaxy known to have high ionization mid-infrared (mid-IR) lines typically associated with active galactic nuclei (AGNs).  Although the faintness of this source precludes a direct spectral analysis, we are able to infer its basic spectral properties using hardness ratios.  X-ray modeling, combined with the nuclear mid-IR characteristics, suggests that NGC 4178 may host a highly absorbed AGN accreting at a high rate with a bolometric luminosity on order of $10^{43}$~ergs~s$^{-1}$.  The black hole mass estimate, based on our {\it Chandra} data and archival VLA data using the most recent fundamental plane relations is $\sim{10^{4}-10^{5}}$~M$_{\sun}$, possibly the lowest mass nuclear black hole currently known.  There are also three off-nuclear sources, two with a similar brightness to the nuclear source at $36\arcsec$ and $32\arcsec$ from the center.  As with the nuclear source, hardness ratios are used to estimate spectra for these two sources, and both are consistent with a simple power-law model with absorption.  These two sources have X-ray luminosities of the order of $\sim{10^{38}}$~ergs~s$^{-1}$, which place them at the threshold between X-ray binaries and ultra-luminous X-ray sources (ULXs).  The third off-nuclear source, located $49\arcsec$ from the center, is the brightest source detected, with an X-ray luminosity of $\sim{10^{40}}$ ergs s$^{-1}$.  Its spectrum is well-fit with an absorbed power law model, suggesting that it is a ULX.  We also fit its spectrum with the Bulk Motion Comptonization (BMC) model and suggest that this source is consistent with an intermediate-mass black hole (IMBH) of mass $(6\pm{2})\times{10^{3}}$~M$_{\sun}$.
\end{abstract}


\section{Introduction}

There is mounting evidence that a significant fraction of supermassive black holes (SMBHs) reside in late-type galaxies, and that a classical bulge is not a requirement for a SMBH to form and grow \citep{filippenko03,barth04,greene04,greene07,satyapal07,dewangen08,ghosh08,mathur08,satyapal08,shields08,barth09,desroches09,gliozzi09,satyapal09,jiang11a,jiang11b,mcalpine11}.  In most late-type galaxies that host an AGN, however, the galaxies have a pseudobulge component, characterized by an exponential surface brightness profile.  And while classical bulges are believed to form through mergers, pseudobulges are thought to form through secular processes~\citep{kormendy04}.  Of the late-type, AGN-hosting galaxies, bulgeless (no evidence even for a pseudobulge) galaxies are by far the rarest.  To date, there are only three such bulgeless disk galaxies that are confirmed to host SMBHs: NGC 4395~\citep{filippenko03,shih03,peterson05}, NGC 1042~\citep{shields08}, and NGC 3621~\citep{satyapal07,barth09,gliozzi09,satyapal09}.  While very large SMBHs ($\gtrsim{10^{6}}$~M$_{\sun}$) likely form through galaxy mergers \citep[e.g.][]{kauffmann00}, leading to a tight correlation between the black hole mass, $M_{\mathrm{BH}}$, and the host galaxy's bulge velocity dispersion $\sigma$~\citep[e.g.]{gebhardt00,ferrarese00,haehnelt02}, it is still unclear how SMBHs form and grow in bulgeless galaxies.  Central to this question is how SMBHs affect, or are affected by, their host galaxy properties.  It has already been shown that the presence and properties of SMBHs do not correlate with galaxy disks or pseudobulges \citep[e.g.][]{kormendy11}.  Interestingly, however, all three bulgeless disk galaxies with SMBHs have nuclear star clusters (NSCs), and there is growing evidence that suggests that the mass of SMBHs and NSCs may be correlated in galaxies that possess both~\citep[e.g.][]{seth08,graham09}.  

Given their rarity, determining the properties of SMBHs in bulgeless disk galaxies is crucial to our understanding of the low end of the SMBH mass function and its relation to host galaxies.  Observationally, the only viable method for finding SMBHs in bulgeless galaxies is through the search for AGNs.  Since bulgeless galaxies are typically dusty, star-forming galaxies, a putative AGN is likely to be missed by optical surveys.  X-ray observations are the ideal tool to search for AGNs in such galaxies since X-rays are generally only produced in the inner nuclear regions of an AGN, and hard X-rays are not substantially affected by absorption.

The goal of this paper is to investigate the X-ray properties of the putative SMBH that lurks at the center of NGC 4178, a bulgeless disk galaxy that was recently found to have prominent mid-IR [NeV] emission associated with the nucleus \citep{satyapal09}.  While the [NeV] emission strongly suggests the presence of an AGN, the size of the {\it Spitzer} InfraRed Spectrograph (IRS) slit ($4.7\arcsec\times{11.3\arcsec}$, at 14.3~\micron) precludes us from confirming the nuclear origin of the emission.  Thus, the presence of a significant X-ray point source counterpart in high-resolution {\it Chandra} data would provide a confirmation of a nuclear SMBH, as X-ray emission associated with starburst activity is generally extended~\citep[e.g.][]{dudik05, gonzalez06, flohic06}.  If it does have a SMBH, NGC 4178 will be only the fourth known truly bulgeless disk galaxy with a SMBH.

NGC 4178 is a highly-inclined ($i\sim{70}\deg$), SB(rs)dm galaxy~\citep{deVaucouleurs91} located within the Virgo Cluster at a distance of 16.8 Mpc~\citep{tully84}.  Based on its nuclear optical spectrum, it is classified as having an HII nucleus \citep{ho97}, and contains an NSC of $\sim{5\times{10^{5}}}$~M$_{\sun}$~\citep{boeker99,satyapal09}.  Other than some asymmetric, locally enhanced H$\alpha$ emission near the outer parts of the disk, the H$\alpha$ distribution of NGC 4178 is typical of that found in star-forming galaxies~\citep{koopmann04}.  The HI distribution is more extended than the optical part of the galaxy and shows no evidence of interactions~\citep{cayatte90}.  Apart from our previous {\it Spitzer} observations, there is no evidence for an AGN in this galaxy.  With these considerations, the presence of a SMBH in this galaxy is highly unexpected.

This paper is structured as follows.  In \S{2}, we describe the {\it Chandra} observations and data reduction, as well as archival VLA data.  We follow with a description of our results, including X-ray spectral modeling in \S{3}.  In \S{4}, we discuss constraints on the nuclear black hole using the bolometric luminosity by presenting an updated $L_{\mathrm{bol}}/L_{\mathrm{[NeV]}}$ correlation.  We compute the Eddington mass and compare our findings with mass estimates from other methods.  In \S{5}, we compare our results to other bulgeless disk galaxies and give a summary and our main conclusions in \S{6}.

\section{Observations and Data Reduction}

\subsection{Chandra Data}

We observed NGC 4178 with {\it Chandra} ACIS-S for 36 ks on 2011 February 19.  The data were processed using \texttt{CIAO v.~4.3} and we retained only events in the energy range $0.2-10$ keV.  We also checked that no background flaring events occurred during the observation.

We used the \texttt{XSPEC v.~12.7.0} software package \citep{arnaud96, dorman01} for the spectral analysis.  For the bright off-nuclear source, we re-binned the spectrum in order to contain at least 15 counts per channel in order to use the $\chi^{2}$ statistic.    To compute the error (90\% confidence) on the flux, we used the \texttt{cflux} model component available in \texttt{XSPEC} as a means to estimate fluxes and errors due to model components \citep{arnaud12}.  For the other sources where low counts ($\lesssim 50$ counts) prevented a direct spectral fit, we employed X-ray hardness ratios as a rough estimator of spectral state.   The hardness ratio we use is defined as: 
\begin{center}
$\frac{\mathrm{hard}}{\mathrm{soft}}\equiv{\frac{\mathrm{counts[\mathrm{2-10~keV}]}}{\mathrm{counts[\mathrm{0.2-2~keV}]}}}$
\vspace{3mm}
\end{center}

\subsection{Archival VLA data}

To search for and constrain the level of radio emission from the detected {\it Chandra} X-ray sources in NGC~4178, we utilized archival {\it Very Large Array} (VLA)\footnote{The National Radio Astronomy Observatory is a facility of the National Science Foundation operated under cooperative agreement by Associated Universities, Inc.} data. The highest angular resolution VLA dataset available was a 560 sec snapshot obtained in B-array at 4.9 GHz ($\sim 1.5''$ beam; program AS314) on Feb 1, 1988, and published by \citet{saikia94}. We used \texttt{AIPS} to calibrate and edit the data using standard procedures.  A final image with beam dimensions $1.95\arcsec\times{1.36\arcsec}$ (position angle = $-12.5\deg$) was produced with the \texttt{IMAGR} task using natural weighting (\texttt{ROBUST} weighting = 2) and 1000 \texttt{CLEAN} iterations.\\

All coordinates listed in this paper refer to the J2000 epoch.

\section{Results}

\begin{figure}
\noindent{\includegraphics[width=8.7cm]{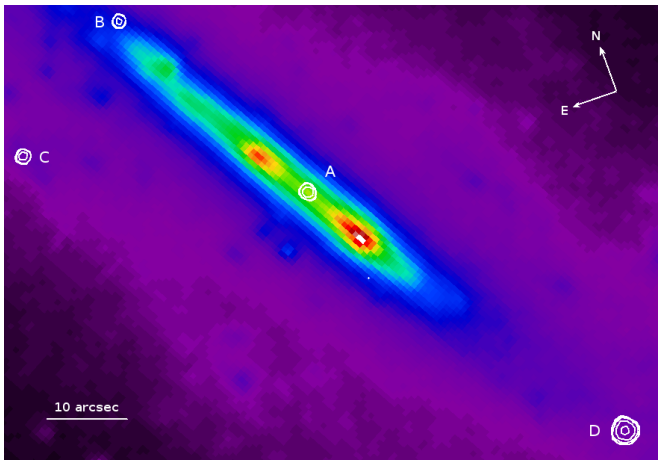}}
\caption{{\it Spitzer} IRAC $3.6~\mu$m image of NGC 4178 from the SPITSOV survey \citep{kenney08} overlaid with contours showing the {\it Chandra} X-ray point sources.  $10\arcsec\simeq{1.6}$ kpc.\\}
\label{chandrasources}
\end{figure}

\begin{figure}
\noindent{\includegraphics[width=8.7cm]{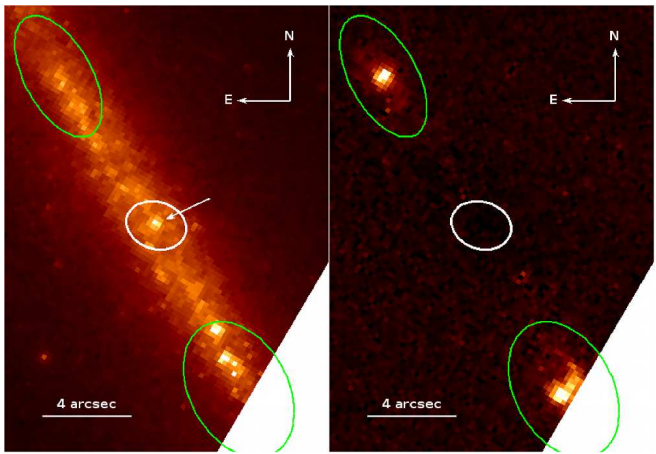}}
\caption{Left: HST NICMOS-3 $H$-band image of NGC 4178 overlaid with the {\it Chandra} nuclear X-ray extraction region in white and the {\it Spitzer} $3.6~\micron$ infrared lobes outlined in green.  Right: HST Pa-$\alpha$ image with the same regions.  The NSC is is marked with an arrow in the $H$-band image.\\}
\label{nucIRXraycontours}
\end{figure}

As can be seen in Figure~\ref{chandrasources}, {\it Chandra} clearly detects a nuclear X-ray source (source A), which appears to be situated symmetrically between two nearly mirrored infrared lobes and is located at RA=$12^{\mathrm{h}}12^{\mathrm{m}}46^{\mathrm{s}}.32$, DEC=$10^{\circ}51\arcmin54\arcsec.61$.  Centroid analysis reveals that the infrared lobes are at the same distance from the nuclear source, about 7.8$\arcsec$, corresponding to a distance of 1.3 kpc at the distance of NGC 4178.  These infrared lobes are likely associated with star formation regions, known from H$\alpha$ studies to be associated with, and confined to, the bar \citep[e.g.][]{martin01}.  This is supported by the presence of Pa-$\alpha$ emission coincident with these lobes (Figure~\ref{nucIRXraycontours}).  Two weaker off-nuclear sources (B and C) are located at RA=$12^{\mathrm{h}}12^{\mathrm{m}}47^{\mathrm{s}}.33$, DEC=$10^{\circ}52\arcmin22\arcsec.52$ 32$\arcsec$ and RA=$12^{\mathrm{h}}12^{\mathrm{m}}48^{\mathrm{s}}.47$, DEC=$10^{\circ}52\arcmin11\arcsec.03$, $35.7\arcsec$ (5.2 kpc) and 36$\arcsec$ (5.9 kpc) from the nuclear source, respectively.  A third, brighter off-nuclear source (D) is located at RA=$12^{\rm{h}}12^{\rm{m}}44^{\rm{s}}.51$, DEC$=10\deg51\arcmin13\arcsec.64$, 49$\arcsec$ (8 kpc) from the nuclear source.  Only one of the off-nuclear sources, source C, appears to have a counterpart in any other band.

In order to determine if the nuclear X-ray source is consistent with the photocenter of the galaxy, the source coordinates were compared to the {\it 2MASS} photocenter.  The {\it 2MASS} photocenter was found to be at RA=$12^{\mathrm{h}}12^{\mathrm{m}}46^{\mathrm{s}}.34$, DEC$=10^{\circ}51\arcmin55\arcsec.1$, $0.6\arcsec\pm{0.6}\arcsec$ off from the X-ray source coordinates.  Thus, we conclude that the nuclear X-ray source is coincident with the {\it 2MASS} photocenter, within the astrometric uncertainties.  After spatially registering the HST $H$-band image with the {\it 2MASS} $H$-band image, we find that the {\it Chandra} source is coincident with the NSC (Figure~\ref{nucIRXraycontours}).

\subsection{The Nuclear Source }

\begin{figure}
\noindent{\includegraphics[width=8.7cm]{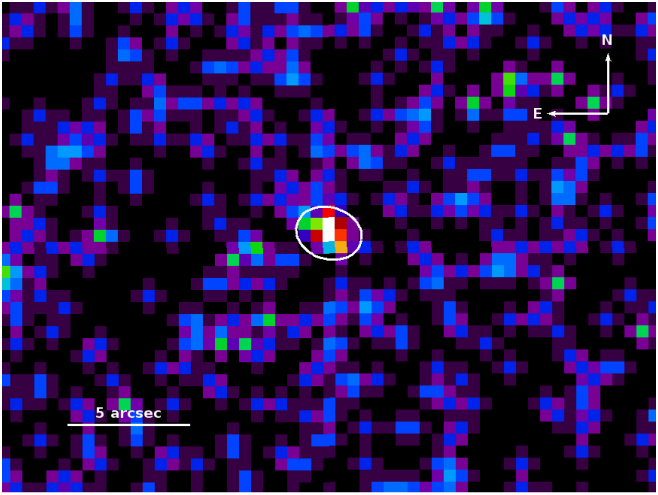}}
\caption{The {\it Chandra} image with the nuclear source region outlined in white.\\}
\label{sourceA}
\end{figure}

The nuclear X-ray source is shown in Figure~\ref{sourceA}.  We detect $37\pm{7}$ ($5.3~\sigma$) X-ray counts ($0.2-10$ keV) from the source.  The nuclear source is strikingly soft, with $31\pm{6}$ counts in the $0.2-2$ keV band and $5\pm{2}$ counts in the $2-10$ keV band (for low counts, Poisson statistics are used to calculate the uncertainty, as described in the approach of \citet{gehrels86}, where the error corresponds to the 84.13\% confidence limit), yielding a hardness ratio of $0.16^{+0.12}_{-0.08}$.  Using a simplified phenomenological power law (PL) model with a Galactic absorption of $1.91\times{10^{20}}$~cm$^{-2}$ \citep[][{\it Chandra} Colden tool)]{dickey90}, and adopting the global intrinsic absorption for NGC 4178 of $N_{H} \simeq 10^{21}$~cm$^{-2}$ \citep{cayatte94}, the observed hardness ratio can be replicated with a photon index $\Gamma=2.6^{+0.6}_{-0.5}$.  Using these parameters, this model predicts an X-ray luminosity of $L_{0.2-10~\mathrm{keV}}=4.6^{+3.6}_{-1.4}\times{10^{38}}$~ergs~s$^{-1}$, and a hard X-ray luminosity is $L_{2-10~\mathrm{keV}}=7.9^{+9.1}_{-4.9}\times{10^{38}}$~ergs~s$^{-1}$, significantly lower than that implied by the observed [NeV] luminosity.\footnote{It should be noted that the apparent luminosity $L_{0.2-10~\mathrm{keV}} = (2.1\pm{0.1})\times{10^{38}}$~ergs~s$^{-1}$ is consistent with the upper limit previously established for NGC 4178 based on {\it Einstein} observations of $L_{0.2-4.0~\mathrm{keV}}\lesssim{2.5\times{10^{40}}}$~ergs~s$^{-1}$ \citep{fabbiano92}.}  Indeed, the bolometric luminosity inferred by the [NeV] luminosity is $\sim{10^{43}}$~ergs~s$^{-1}$ (see \S\ref{sec:boldiscuss}), 5 orders of magnitude higher than the hard X-ray luminosity predicted by this simplified phenomenological PL model.  The bolometric correction factor is $\kappa_{2-10~\rm{keV}}\sim{10^{5}}$, much too high for any realistic AGN spectral energy distribution (SED)~\citep{vasudevan09}.\footnote{In predicting the bolometric luminosity, we included all of the [NeV] $14.3~\micron$ flux.  In principle, [NeV] emission can be produced by shocked gas associated with a starburst-driven superwind~\citep[e.g.][]{veilleux05}.  However, in the case of NGC 4178, the absence of notable Pa-$\alpha$ emission near the photocenter (Figure~\ref{nucIRXraycontours}) suggests that there is no vigorous star formation around the nuclear X-ray source.  Moreover, the optical spectrum is typical of an HII region and is not consistent with shocks~\citep{ho97}.}

We therefore postulate that there is highly localized absorption around the central source.  Indeed, analysis of data from {\it Chandra} Deep Field North has shown that more complex spectra are characteristic of AGNs in local galaxies due to the ubiquity of heavy absorption, and a simple PL cannot reliably be estimated from hardness ratios~\citep{brightman12}.  With a covering fraction of $0.99$ and absorption $N_{H}=5\times{10^{24}}$~cm$^{-2}$, a partially-absorbed scenario combined with a simple PL can account for the hardness ratio with $\Gamma = 2.3^{+0.6}_{-0.5}$.  With these parameters, the intrinsic X-ray luminosity is $L_{\mathrm{0.2-10~keV}}=3.1^{+1.5}_{-0.5}\times{10^{40}}$~ergs~s$^{-1}$ and the hard X-ray luminosity is $L_{\mathrm{2-10~keV}}=8.6^{+10.4}_{-5.5}\times{10^{39}}$~ergs~s$^{-1}$.  The partially-absorbed scenario can be physically interpreted as a strongly-accreting black hole imbedded in a heavy medium in which a few holes have been ``punched out" by strong X-ray flux.  This is in line with our understanding of black hole accretion, our understanding of the interstellar environment of late-type spiral galaxies, and it is consistent with the bolometric luminosity.

We point out that the observed X-ray luminosity is low, and therefore by itself we cannot exclude the possibility that it is produced by X-ray binaries in the nuclear star cluster.  However, the [NeV] luminosity is $8.23\times{10^{37}}$~ergs~s$^{-1}$, which is at the the low end of the range observed in standard optically-identified AGNs~\citep{pereira-santaella10}.  Indeed, the [NeV] luminosity in NGC 4178 exceeds that of NGC 3621, which has recently been confirmed to have an optical Seyfert 2 spectrum using high resolution Keck observations~\citep{barth09}.  Moreover, the bolometric luminosity implied by the [NeV] luminosity is two orders of magnitude greater than that of NGC 4395, which is indisputably an AGN~\citep{filippenko03} (see~\ref{sec:boldiscuss}).  Therefore, the [NeV] luminosity, combined with our X-ray results, strongly suggests that the X-ray source is due to an AGN.

\subsection{Off-Nuclear Sources}

\subsubsection{Source B}

We detected $20\pm{5}$ net counts in the $0.2-10$ keV range for source B.  The hardness ratio is $0.41^{+0.38}_{-0.22}$.  At a radial distance of 31.6$\arcsec$, the known intrinsic absorption for NGC 4178 is $N_{H} \simeq 10^{21}$~cm$^{-2}$ \citep{cayatte94}.  With this absorption, source B's hardness ratio can be reproduced by an absorbed PL model with $\Gamma = 1.7^{+0.7}_{-0.6}$.  The corresponding X-ray luminosity is $L_{0.2-10~\mathrm{keV}} = 2.2^{+1.5}_{-0.5}\times{10^{38}}$~ergs~s$^{-1}$.  In the hard X-ray band, the corresponding luminosity is $L_{2-10~\mathrm{keV}} = 1.2^{+1.7}_{-0.8}\times{10^{38}}$~ergs~s$^{-1}$.

\subsubsection{Source C}

The net counts in the $0.2-10$ keV range for source C are $26\pm{5}$, with a hardness ratio of $0.20^{+0.19}_{-0.12}$.  This relatively low hardness ratio can be reproduced by an absorbed PL model with $\Gamma = 2.3^{+0.9}_{-0.6}$ and $N_{H} = 10^{21}$~cm$^{-2}$, giving an X-ray luminosity of $L_{0.2-10~\mathrm{keV}}=6.2^{+6.8}_{-1.2}\times{10^{38}}$~ergs~s$^{-1}$ and a hard X-ray luminosity of $L_{2-10~\mathrm{keV}}=1.7^{+2.4}_{-1.2}\times{10^{38}}$~ergs~s$^{-1}$.

\begin{figure}
\noindent{\includegraphics[width=8.7cm]{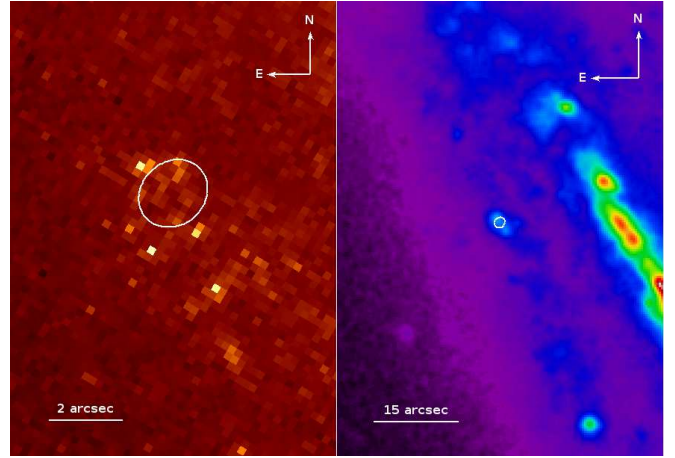}}
\caption{Left: HST NICMOS-3 H-band image of NGC 4178 overlaid with the \texttt{celldetect} X-ray extraction region for source C.  Right: Smoothed SDSS g-band image with the same contour.  $1\arcsec\simeq{160}$ parsecs.\\}
\label{hstcontours}
\end{figure}

\subsubsection{Source D}

Our brightest X-ray source, source D, had $575\pm{24}$ total counts in the $0.2-10$ keV range, enough to directly fit a spectrum.  We tested the variability of this source by extracting the light curve in the $0.2-10$ keV range and applying a $\chi^2$ test, and found that the source flux was likely constant during the observation ($\chi^{2}/\mathrm{dof}= 22.584/39$, $P_{\chi^{2}}=0.98$).

The spectrum is well fit ($\chi^{2}_{\mathrm{red}}=0.997$ for 33 degrees of freedom) using an absorbed PL model with $\Gamma=1.24\pm{0.12}$ and setting the intrinsic absorption $N_{H}=10^{21}$~cm$^{-2}$.  This yields an X-ray luminosity of $L_{\mathrm{0.2-10~keV}}=(7.9\pm{0.8})\times{10^{39}}$~ergs~s$^{-1}$ and a hard X-ray luminosity of $L_{\mathrm{2-10~keV}}=(5.9\pm{0.9})\times{10^{39}}$~ergs~s$^{-1}$.  If we allow the intrinsic absorption to vary, a comparable fit ($\chi^{2}_{\mathrm{red}}=1.02$ for 32 degrees of freedom) is achieved with $N_{H}=1.5\pm{0.14}\times{10^{20}}$ cm$^{2}$ and $\Gamma=1.31\pm{0.23}$, yielding an intrinsic X-ray luminosity $L_{\mathrm{0.2-10~keV}}=(8.0\pm{1.0})\times{10^{39}}$~ergs~s$^{-1}$ and a hard X-ray luminosity of $L_{\mathrm{2-10~keV}}=(5.8\pm{0.7})\times{10^{39}}$~ergs~s$^{-1}$.

This large luminosity gives us a second method of spectrally characterizing the source.  Recent work has shown that the physically motivated Bulk Motion Comptonization (BMC) model may provide an effective means of estimating the mass of accreting black holes~\citep{gliozzi09,shaposhnikov09,gliozzi11}.  This technique relies on the self-similarity of black holes and their accretion characteristics to relate the photon index $\Gamma$ with the BMC model normalization $N_{\rm{BMC}}$.  In short, the BMC model convolves the inverse Comptonization of X-ray photons by thermalized electrons with the inverse Comptonization of X-ray photons by electrons with bulk relativistic motion~\citep[see][for details on the BMC model]{titarchuk97}.  The BMC model has 4 free parameters: the temperature $kT$, the spectral index $\alpha$, related to the photon index $\Gamma$ by $\alpha = \Gamma - 1$, a parameter $\log{A}$ related to the fraction of Comptonized seed photons $f$ by $A = f/(f - 1)$, and the model normalization $N_{\rm{BMC}}$.

\begin{figure}
\noindent{\includegraphics[width=8.7cm]{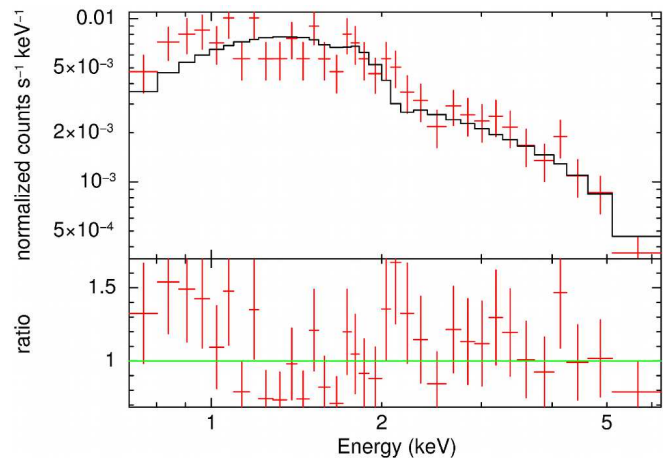}}
\caption{Spectrum of source D and data/model ratio to a BMC model with galactic absorption.\\}
\label{srcDspectrum}
\end{figure}

The BMC model gives an excellent fit ($\chi^{2}_{\mathrm{red}}=1.00$ for 31 degrees of freedom) with the intrinsic absorption set to $N_{H}=5\times{10^{20}}$ cm$^{2}$, $kT=0.48^{+0.08}_{-0.07}$, $\alpha=0.77^{+0.44}_{-0.25}$, and $\log{A}=7.1$ (Figure~\ref{srcDspectrum}).  The normalization is $N_{\mathrm{BMC}}=1.39\times{10^{-6}}$.  The corresponding X-ray luminosity is is $L_{0.2-10~\mathrm{keV}}=(6.5\pm{0.4})\times{10^{39}}$~ergs~s$^{-1}$, and the hard X-ray luminosity is $L_{2-10~\mathrm{keV}}=(5.1\pm{0.5})\times{10^{39}}$~ergs~s$^{-1}$.  Five different spectral patterns of Galactic black hole systems with mass and distance well constrained are provided in \citet{gliozzi11}.  These are the reference sources utilized in the X-ray scaling method.  To be conservative in the estimate of the black hole mass of Source C, we have used all five reference patterns and computed the average and standard deviation of the five $M_{\rm{BH}}$ values, resulting in $\langle{M_{\rm{BH}}}\rangle=(6\pm{2})\times{10^{3}}$~M$_{\sun}$.

\subsubsection{The Nature of The Off-Nuclear Sources}
With the exception of source C, the lack of noticeable counterparts in other bands (see Figure~\ref{chandrasources}) suggests that these sources are not foreground objects.  Source C is coincident with an extended ($\sim$ 800 $\times$ 520 parsecs) region (Figure~\ref{hstcontours}), especially evident in high-resolution HST data.  This region appears to be an area of extensive star formation, though it is not associated  with significant Pa-$\alpha$ emission.  A priori, we cannot rule out that these sources are background AGN that appear to be located within NGC 4178 by chance.  With the known hydrogen column density of $\sim{10^{21}}$~cm$^{-2}$, a background X-ray source would not be heavily absorbed by NGC 4178, even in the soft band.  Using results from the {\it Chandra} Deep Field-South (CDF-S), we can calculate the likely number of hard X-ray sources in our field ($\sim{4\times{10^{-4}}}$ deg$^{-2}$) for a given flux that could be expected to occur by chance~\citep[e.g.][]{tozzi01}.  For sources B and C, with hard X-ray fluxes on order of $\sim{10^{-15}}$~ergs~s$^{-1}$~cm$^{-2}$, the expected number of sources is $\sim{5}$, so sources B and C may indeed be background objects.  However, for source D, with hard X-ray flux on order of $\sim{10^{-13}}$~ergs~s$^{-1}$~cm$^{-2}$, the expected number of sources is $\sim{0.1}$.  We therefore conclude that source D is likely local to NGC 4178.  Source D has a luminosity of $\sim{10^{40}}$~ergs~s$^{-1}$, and is therefore consistent with a ULX~\citep[e.g.][]{swartz04,winter06,berghea08}.  Source D's spectrum is also well fit by the BMC model, suggesting that it may be an IMBH.   VLA observations by \citet{cayatte90} of neutral hydrogen in NGC 4178 show a heightened concentration ($N_{H}=1.48\times{10^{22}}$~cm$^{-2}$) of neutral hydrogen coincident with source D.~~\citet{niklas95} gives the position of this source as $\mathrm{RA}\simeq{12^{\mathrm{h}}12^{\mathrm{m}}43^{\mathrm{s}}}$, $\mathrm{DEC}\simeq{10^{\circ}50\arcmin59\arcsec}$.  This concentration is very diffuse, however, so it is uncertain that it is related to source D.

\section{Constraints on the Nuclear Black Hole Mass}
\subsection{Bolometric Luminosity and Eddington Mass Limit of the AGN}
\label{sec:boldiscuss}

\begin{figure}
\noindent{\includegraphics[width=8.7cm]{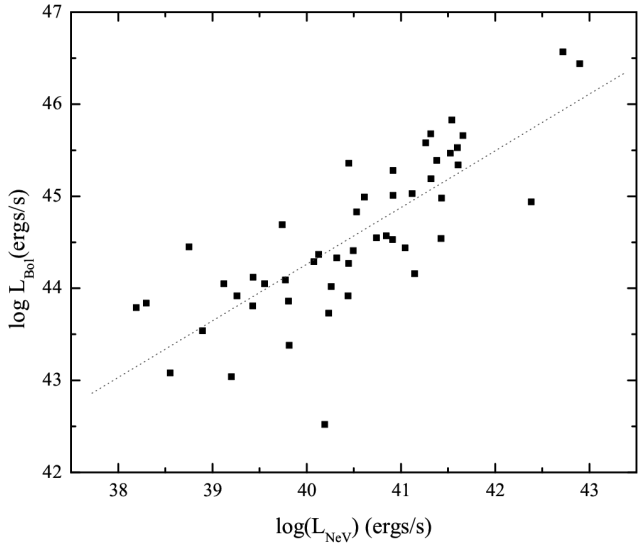}}
\caption{AGN bolometric luminosity as a function of [NeV] $14.32~\micron$ luminosity.  It is clear that the [NeV] luminosity is strongly correlated with the bolometric luminosity over a wide range of luminosities.\\}
\label{lbolnevplot}
\end{figure}

Our IR and X-ray observations allow us to derive a mass estimate for the nuclear black hole.  In our previous work, we showed that the [NeV] luminosity is tightly correlated with the bolometric luminosity of the AGN in a sample of optically identified AGNs.  We can therefore use the [NeV] luminosity to obtain an estimate of the Eddington mass of the black hole~\citep{satyapal07}, and therefore set a lower mass limit on the AGN.  Since the publication of our previous work, there have been a number of additional mid-IR [NeV] fluxes available in the literature.  We therefore update our previously published relation between the [NeV] $14.32~\micron$ luminosity and the AGN bolometric luminosity using the most up-to-date mid-IR line fluxes of AGNs observed by {\it Spitzer} \citep{haas05, weedman05, ogle06, dudik07, dudik09, gorjian07, cleary07, armus07, deo07, tommasin08, tommasin10, dale09, veilleux09, pereira-santaella10} that have well-characterized nuclear SEDs.  In Figure~\ref{lbolnevplot}, we plot $L_{\mathrm{bol}}$ versus $L_{\mathrm{[NeV]}}$, which shows a strong correlation.  The Spearman rank correlation coefficient is 0.83, with a probability of chance correlation of $10^{-13}$.  The bolometric luminosities for this updated sample ranged from $\sim{5\times{10^{42}}}$~ergs~s$^{-1}$ to $\sim{5\times{10^{46}}}$~ergs~s$^{-1}$ and the black hole masses ranged from $\log{M_{\mathrm{BH}}}=6.15$ to $\log{M_{\mathrm{BH}}}=9.56$.  The best-fit linear relation yields:
\begin{center}
$\log{L_{\mathrm{bol}}}=0.615\log{L_{\mathrm{[NeV]}}}+19.647$~ergs~s$^{-1}$ 
\end{center}
with an RMS scatter of 0.53 dex.  Using the known [NeV] $14.32~\micron$ nuclear luminosity for NGC 4178 of $8.23\times{10^{37}}$~ergs~s$^{-1}$ \citep{satyapal09}, the predicted nuclear bolometric luminosity of the AGN is $L_{\mathrm{bol}}={9.2^{+22}_{-6.5}\times{10^{42}}}$~ergs~s$^{-1}$.  The Eddington mass limit for the nuclear black hole in NGC 4178 is then $M_{\rm{BH}}\geq{7.1^{+17}_{-5.0}}\times{10^{4}}$~M$_{\sun}$.

\subsection{Archival VLA Data and The Fundamental Plane}

At the positions of the four Chandra sources, we did not find any significant emission in the VLA 4.9 GHz image and measured $3~\sigma$ point source limits of $<0.23$ (A), $<0.17$ (B), $<0.15$ (C), and $<0.16$ (D) mJy.  These limits are consistent with the original analysis of~\citet{saikia94} where no significant radio sources were detected.  Interestingly, we note a local radio maximum in the VLA image that is $1.7\arcsec$ to the northeast of the central Chandra source (A) with a position, RA=$12^{\rm{h}}12^{\rm{m}}46^{\rm{s}}.40$, DEC$=10\deg51\arcmin55\arcsec.9$. The peak of 0.184 mJy/beam is 3.4 times the average rms noise measured in adjacent background regions. At 4.9 GHz, the single-dish flux of NGC~4178 is 12 mJy \citep{vollmer04} and this peak could easily be an artifact due to unmodelled diffuse radio emission from the galaxy.   The only other VLA datasets available in the archive were obtained at 1.4 GHz, but these lower-resolution data are dominated by the diffuse star forming regions seen in the Effelsberg data and the VLA 1.4 GHz D-array image of \citet{condon87}.

If we use the radio luminosity as an upper limit, we can use the hard X-ray luminosity to estimate the upper limit on the nuclear black hole mass through the so-called fundamental plane \citep{merloni03,falcke04,merloni06,gultekin09}, assuming that the relation extends to lower luminosities and black hole masses.  Assuming a 5 GHz flux density for the nucleus of 0.2 mJy, the X-ray and radio luminosity would imply a nuclear black hole mass of $\sim{2.0^{+8.2}_{-1.6}\times{10^{5}}}$~M$_{\sun}$ (the RMS scatter is $\sim{0.7}$), consistent with the Eddington mass limit derived in the previous section.\footnote{In calculating this mass, we have used the hard X-ray luminosity of $8.6^{+10.4}_{-5.5}\times{10^{39}}$~ergs~s$^{-1}$ from the heavily absorbed scenario.  A lower X-ray luminosity would actually {\it increase} the upper limit to the black hole mass, according to the fundamental plane.} The assumed radio flux value corresponds to the local peak seen in the VLA image in close proximity to the X-ray nuclear source, and is consistent with the strict upper limit measured at the X-ray position.

\subsection{Other Considerations}
\label{sec:otherconsid}

The presence of an NSC in NGC 4178 provides another estimate of the mass of the central black hole.  \citet{satyapal09} estimated the mass of the NSC to be $\sim{5\times{10^5}}$~M$_{\sun}$, similar to the NSC mass in NGC 4395.  In cases with a known nuclear star cluster and black hole masses, the ratio of the black hole mass to nuclear star cluster mass, $M_{\mathrm{BH}}/M_{\mathrm{NSC}}$, generally ranges from $0.1-1$~\citep{seth08,graham09}.  The lower limit on the black hole mass is then $\sim{5\times{10^4}}$~M$_{\sun}$, consistent with the lower limit set by the Eddington mass (although some galaxies have a BH/NSC mass much less than 0.1, such as M33), and the upper limit on the black hole mass is $\sim{5\times{10^{5}}}$~M$_{\sun}$, consistent with the upper limit implied by the radio luminosity.

With the heavily obscured scenario, we can further constrain the black hole mass.  Assuming that the NGC 4178 has an X-ray luminosity of $\sim{10^{40}}$~ergs~s$^{-1}$, the bolometric correction factor becomes $\kappa_{\mathrm{2-10~keV}}\simeq{10^3}$.  This finding can in turn be used to constrain the black hole mass by exploiting the correlation between $\kappa_{\rm{2-10~keV}}$ and $L_{\mathrm{bol}}/L_{\mathrm{Edd}}$~\citep{vasudevan09}.  Using the correlation that for systems with very high bolometric correction factors, $L_{\mathrm{bol}}/L_{\mathrm{Edd}}\gtrsim{0.2}$, we obtain $M_{\mathrm{BH}}\simeq{3.8\times{10^{5}}}$~M$_{\sun}$, consistent with the upper mass limit implied by the radio luminosity.

\section{Comparison to AGN in Other Late-Type, Bulgeless Galaxies}

With a solid X-ray source detection at the center of NGC 4178, we can add this galaxy to the growing collection of bulgeless, extremely late-type disk galaxies with confirmed AGN.  The best-studied definitively bulgeless disk galaxy with an AGN is NGC 4395, which shows the hallmark signatures of a type 1 AGN \citep[e.g.][]{filippenko03,lira99,moran99}.  The bolometric luminosity of the AGN is $\sim{10^{40}}$~ergs~s$^{-1}$ \citep{filippenko03}, nearly three orders of magnitude lower than the estimated bolometric luminosity of the AGN in NGC 4178.  The black hole mass of NGC 4395, determined by reverberation mapping, is $M_{\mathrm{BH}}=(3.6\pm{1.1})\times{10^{5}}$~M$_{\sun}$, and does not appear to be radiating at a high Eddington ratio \citep{peterson05}.  The bolometric luminosity of the AGN in NGC 1042, as estimated from H$\alpha$ measurements, is $\sim{8\times{10^{39}}}$~ergs~s$^{-1}$, and the central black hole is estimated to be between 60~M$_{\sun}$ and $3\times{10^{6}}$~M$_{\sun}$~\citep{shields08}.  With the updated $L_{\mathrm{bol}}/L_{\mathrm{[NeV]}}$ relationship, the estimated bolometric luminosity of the AGN in NGC 3621 is~$6.8^{+16}_{-4.8}\times{10^{42}}$~ergs~s$^{-1}$ and the Eddington mass is $5.2^{+12}_{-3.7}\times{10^{4}}$~M$_{\sun}$, making NGC 4178 and NGC 3621 the most luminous AGNs in extremely late-type galaxies currently known.    The likelihood that the AGN in NGC 4178 is heavily absorbed, combined with the high photon index, suggests that it is accreting at a high rate.  The black hole mass of the AGN in NGC 4178 is $\sim{10^{4}-10^{5}}$~M$_{\sun}$, possibly lower than the black hole in NGC 4395.

\section{Summary and Conclusions}

We have analyzed the X-ray characteristics of NGC 4178 from a 36 ks {\it Chandra} observation.  The X-ray data, combined with considerations from the mid-IR and radio properties of the galaxy, have led us to the following results:

\begin{enumerate}
\item{There is a faint but statistically significant ($5.3~\sigma$), unresolved X-ray source at the center of NGC 4178, confirming the presence of an AGN.  The hardness ratio gives some clues about the spectral state, which is consistent with a scenario where the source is accreting at a high rate with $\Gamma\simeq{2.3}$.  The softness of this source, combined with the discrepancy between the [NeV] luminosity and the observed X-ray luminosity, supports the scenario where NGC 4178 hosts an AGN embedded in a heavy absorber, accreting at a high rate.}
\item{The bolometric luminosity of the AGN in NGC 4178 predicted by our mid-IR results is $9.2\times{10^{42}}$~ergs~s$^{-1}$, significantly higher than that found in any other extremely late-type, bulgeless disk galaxy.}
\item{The updated bolometric luminosity, combined with other lines of evidence such as the fundamental plane and the correlation between the mass of nuclear star clusters and their resident SMBHs, have led us to conclude that the AGN in NGC 4178 is powered by a black hole of $\sim{10^{4}-10^{5}}$~M$_{\sun}$.}
\item{Two weak off-nuclear sources found in NGC 4178 have X-ray luminosities consistent with very bright XRBs or ULXs, although we cannot rule out the possibility that they are background objects.  A third off-nuclear source is very bright and was directly fit with a PL model, showing that it is a ULX located in NGC 4178 about 8 kpc from the nuclear source.  It was also directly fit with a BMC model, suggesting that it may be an IMBH of $\sim{6\times{10^{3}}}$~M$_{\sun}$.}
\end{enumerate}

\section{Acknowledgements}

It is a pleasure to thank Tim Jordan for his invaluable help carefully compiling the fluxes used to construct Figure~\ref{lbolnevplot}.  We thank the referee for their very careful review and insightful comments that significantly improved the paper.  This research has made use of the NASA/IPAC Extragalactic Database (NED), which is operated by the Jet Propulsion Laboratory, California Institute of Technology, under contract with the National Aeronautics and Space Administration.  N.~S.~and S.~S.~gratefully acknowledge support by the {\it Chandra} Guest Investigator Program under NASA grant G01-12126X.  Work by C.~C.~C.~at NRL is supported in part by NASA DPR S-15633-Y.
\\
\\

\end{document}